\documentclass[lettersize,journal]{IEEEtran}
\usepackage{amsmath,amsfonts,amssymb}
\usepackage{algorithmic}
\usepackage{algorithm}
\usepackage{array}

\usepackage[caption=false,font=normalsize,labelfont=sf,textfont=sf]{subfig}
\usepackage{textcomp}
\usepackage{stfloats}
\usepackage{url}
\usepackage{verbatim}
\usepackage{graphicx}
\usepackage{cite}
\usepackage{amsfonts, amsmath, amssymb, amsthm}
\usepackage{indentfirst}
\usepackage{xcolor}
\usepackage{array}
\usepackage[all]{xy}

\newcommand{\Id}{\mathrm{Id}}

\newcommand{\ima}{\mathop{\mathrm{Im}}}

\newcounter{forterm}
\newenvironment{term}
{\refstepcounter{forterm} \par \medskip \textbf{Definition \arabic{forterm}.} \it } %
{}

\newcounter{forprop}
\newenvironment{prop} %
{\refstepcounter{forprop} \par \medskip \textbf{Proposition \arabic{forprop}.} \it } %
{}

\newtheorem{manualtheoreminner}{Theorem}
\newenvironment{manualtheorem}[1]{%
  \renewcommand\themanualtheoreminner{#1}%
  \manualtheoreminner
}{\endmanualtheoreminner}

\newenvironment{myproof}{\par \medskip \textbf{Proof. }}{\hfill$\square$}
\newcounter{forpic}

\begin{document}

\title{On the Interference Cancellation by Reduced Channel Zero Forcing Class of Precodings in Massive MIMO Systems}

\author{Dmitry~Mineev,
      Evgeny~Bobrov,
       and  Viktor~Kuznetsov%
%         and~Danila~Zaev% <-this % stops a space

\thanks{The work has been supported by Huawei Technologies. Evgeny~Bobrov is with Huawei Technologies, Russian Research Institute, Moscow Research Center, and M.V. Lomonosov Moscow State University, Russia (eugenbobrov@ya.ru). Dmitry~Mineev and Viktor~Kuznetsov are with Huawei Technologies, Russian Research Institute, Moscow Research Center (mineyev.dmitry@huawei.com, kuznetsov.victor@huawei.com).}} %<-this  stops a space

%\thanks{Manuscript received April 19, 2021; revised August 16, 2021.}
\markboth{Journal of \LaTeX\ Class Files,~Vol.~14, No.~8, August~2021}%
{Shell \MakeLowercase{\textit{et al.}}: A Sample Article Using IEEEtran.cls for IEEE Journals}

%dealing with the ever-increasing problem of inter-cell interference.
%Advanced receivers and transmitters are a key component of the 5th Generation (5G) wireless systems
%In order to suppress the residual interference in multiuser multiple-input and multiple-output (MU-MIMO) systems, a number of low-complexity interference-aware types of receivers have been considered. 

\maketitle

\begin{abstract}
In this paper, we study the interference cancellation capabilities of receivers and transmitters in multiple-input-multiple-output (MIMO) systems using theoretical calculations and numerical simulations in Quadriga. We study so-called Reduced Channel Zero-Forcing (RCZF) class of precoding as well as Minimum MSE Interference Rejection Combiner (MMSE-IRC) and QR Maximum Likelihood Detection (QR-MLD) receivers. Based on very simple but extremely useful algebraic manipulations, their asymptotical equivalence is proven analytically and demonstrated via simulations. Our theoretical and experimental results confirm that MMSE-IRC and QR-MLD receivers in combination with the RCZF precoding provide complete interference suppression asymptotically.


\end{abstract}

\section{Introduction}
Massive multi-input multi-output (MIMO) systems are considered to be key component of fifth-generation (5G) networks~\cite{RobustIRC, MIMONextGeneration, DetectionEquivalence, bobrov2021massive, Bjornson}, because they are designed to serve many more users with a much higher overall bandwidth~\cite{HighPerformanceDetection}. It has been proved~\cite{Architecture5G, Tutorial5G} that a massive MIMO system can drastically increase the spectral efficiency of the wireless channel. One of the reasons for this comes from the fact that, as the number of transmitting antennae at the base station (BS) grows, both inter-cellular and intra-cellular interference can be effectively negated~\cite{Bjornson_tb_17}.

The first layer of MIMO system containing scheduling procedure also addresses the phenomenon of interference, and many useful algorithms of user selection and resource allocation for the MIMO downlink have been investigated in~\cite{MISO_RZF} (see also references therein). Scheduling techniques try to minimize the interference taking into account geometrical alignment of users, but the residual inter-user interference is still significant. For proper detection at each user equipment (UE), this residual interference must be suppressed by means of the second layer of MIMO system. Furthermore, the first layer of MIMO system provides the decision for rank selection of every UE, and it is up to the second layer again to devise precoding with given user ranks.

A lot of effort has been put into developing precoding and detection methods for multi-user MIMO systems~\cite{Bjornson_tb_17, 50Years, IterativeDetection, MultiuserDetection, bobrov2021adaptive, bobrov2022lbfgs}. In this paper we introduce the novel class of precoding called Reduced Channel Zero-Forcing (RCZF) and study its properties. RCZF precoding generalizes known constructions such as EZF~\cite{SVD}. In Section~\ref{sec:system_model} we define necessary terms and provide two descriptions of MIMO system suitable for our study. In Section~\ref{sec:rczf_precoding} we proceed to prove Theorem~\ref{theorem:rczf} which states that RCZF class precoding is capable of eliminating interference while others are not. In Section~\ref{sec:detection_options} we provide detailed description of two widespread detection schemes, namely Minimum MSE Interference Rejection Combiner (MMSE-IRC)~\cite{IRC, PotentialIRC, ExperimentalIRC} and QR Maximum Likelihood Detection (QR-MLD)~\cite{MLD}. Theoretical part of this paper culminates in Theorems~\ref{theorem:mmse_irc} and~\ref{theorem:qr_mld_2} stating that these two detection schemes suppress interference combined with RCZF precoding. In Section~\ref{sec:experiments} we support all the claims of the paper with Quadriga~\cite{Quadriga} based experiments and present their description and graphical evidence. All necessary algebraic notations are given in Tab~\ref{tab:notations}.
\IEEEpubidadjcol

\section{System Model}\label{sec:system_model}

The multi-user multi-input multi-ouput (MU-MIMO) model is described by the following linear system, where $r_k \in \mathbb{C}^{p_k}$ is a vector of detected symbols at the $k$-th receiver, $s \in \mathbb{C}^p$ is a common vector of sent symbols, $H_k \in \mathbb{C}^{q_k \times t}$ is a channel matrix of the $k$-th receiver, $W \in \mathbb{C}^{t \times p}$ is a common precoding matrix, $G_k \in \mathbb{C}^{p_k \times q_k}$ is a detection matrix at the $k$-th receiver, and $n_k \sim \mathcal{CN}(0,I_{p_k})$ is a $k$-th receiver noise:
 \begin{equation}\label{System Model}
  r_k = G_k(H_kWs+n_k).
 \end{equation}

In the MIMO system, it is possible to send several information symbols to a multi-antenna user on a single physical resource. The number of such symbols is called the rank of the $k$-th user and is denoted by $p_k$, and $p_1 + \dots + p_l =p$ means the total number of layers. The constant $t$ means number of transmitting antennas on the base station, $q_k$ means number of receiving antennas of the $k$-th user, where $q_1 + \dots + q_l = q$. We assume they are related as follows: $p_k \leqslant q_k \leqslant t$. 

\begin{figure}
    \centering
    \xymatrix@R=1em@C=6.3em{
    S_1 & \ar@{-->}[r]|-{+n_1} & U_1 \ar[r]|-{G_1} & S_1 \\
    \vdots & & \vdots & \vdots \\
    \bigoplus \ar[r]|-{\oplus W_k}  & A \ar[uur]|-{H_1} \ar[r]|-{H_k}  \ar[ddr]|-{H_l} & U_k \ar[r]|-{G_k}  & S_k \\
    \vdots & & \vdots & \vdots \\
    S_l & \ar@{-->}[r]|-{+n_l} & U_l \ar[r]|-{G_l} & S_l}
    \caption{Spaces and maps description of MIMO system.}
    \label{fig:algebraic_system}
\end{figure}
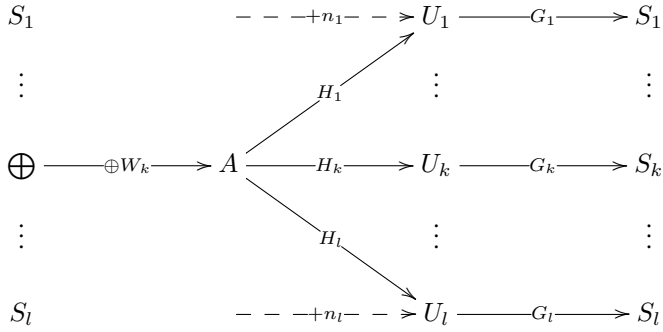

For our purposes, the following equivalent description of MIMO system will be convenient (see Fig.~\ref{fig:algebraic_system}). The noiseless linear MIMO system consists of three sets of complex vector spaces: $S_k$ --- \textit{symbol spaces}, $A$ --- \textit{antenna space}, and $U_k$ --- \textit{user spaces}; and of three sets of linear maps between them: $W_k \colon S_k \to A$ --- \textit{precoding}, $H_k \colon A \to U_k$ --- \textit{channel}, and $G_k \colon U_k \to S_k$ --- \textit{detection}.

Note that precoding maps separate symbol spaces to the common antenna space, channel maps common antenna space to separate user spaces, and detection is completely independent for each user, mapping separate user spaces to separate symbol spaces. We also assume that channel maps $H_k$ have full rank, i.e., $\mathop{\mathrm{rk}}H_k=\dim U_k = q_k$.
\par Noisy MIMO system is obtained by adding randomly distributed noise vectors $n_k \in U_k$. Roughly speaking, the goal of MIMO system is to make $G_kH_kW_k$ as close as possible to identity map (this is an effective signal addressed to the user) and to make $G_kH_kW_j$ for $j \neq k$ as close as possible to zero map (this is interference with the signal intended for other users). \par
\begin{term}
We say that noiseless MIMO system \textbf{cancels interference} if $G_kH_kW_k = \Id$ and $G_kH_kW_j=0$ for $j\neq k$.
\end{term}
\par All vector spaces participating in MIMO system are equipped with natural bases: for symbol spaces, the coordinates in the base are the symbols themselves; for antenna and user spaces the coordinates in the base are values on antennae (either transmitted or received). We will denote matrices of linear maps in MIMO system in these bases by the same letters $W,H,G$. This notation coincides with the standard matrix description of MIMO system.
\par

\section{RCZF Precoding}\label{sec:rczf_precoding}
\begin{term}\label{def:RCZF}
Set of precoding maps $W_k$ is called \textbf{Reduced Channel Zero-Forcing} (abbreviated as RCZF) if there exist matrices $V_k$ such that
\begin{itemize}
    \item $V_k = B_k H_k$ for some $B_k$;
    \item $V_k W_j = 0$ for $j \neq k$;
    \item $\mathop{\mathrm{rk}} V_kW_k = \mathop{\mathrm{rk}}W_k=\dim S_k=p_k$.
\end{itemize}
\end{term}

We will refer to matrices $V_k$ as \textit{reduced channel} matrices. In particular, usual zero-forcing is obtained if $B_k = I$, and $V_k=H_k$ are whole channel matrices. Less trivial example is given by eigen zero-forcing (abbreviated as EZF,~\cite{SVD}), which determines $B_k$ by the following procedure. Write SVD decomposition of channel $H_k = \mathcal{U}_k\mathcal{S}_k\mathcal{V}_k$. Then $B_k = I' \mathcal{S}_k^{-1} \mathcal{U}_k^{-1}$, where $I' = \begin{pmatrix} I | 0 \end{pmatrix}$, so that $V_k$ is the upper cut of $\mathcal{V}_k$. Precoding itself is calculated as pseudo-inverse of concatenated matrices $V_k$, which is one of possible ways to satisfy conditions of RCZF. \par
\par Let us make another important remark on the definition of RCZF. Instead of usual zero-forcing relations such as $H_kW_k=I$ or, more realistically, $H_kW_k$ is non-zero diagonal, we make a weaker requirement on the full rank. This is done for the general case when the number of layers for the user $k$ (i.e., $\dim S_k=p_k$), is strictly less than $\dim V_k$ (i.e., the dimension of the reduced channel). This happens, for instance, if the number of layers is strictly less than the number of user antennae (i.e., $\dim U_k=q_k$), and we do full zero-forcing with $V_k=H_k$.

\begin{table}
\centering
\small
\resizebox{\columnwidth}{!}{%
\begin{tabular}{  c | l  }
$H_k$   & $t\times q_k$  (64$\times$4) channel matrix of user $k$ \\
$V_k$     & $t \times p_k$ (64$\times$2) layers matrix of UE $k$ \\
$B_k$ & $p_k \times q_k$ (4$\times$2) reducing channel to layers, i.e., $H_k B_k = V_k$ \\
$W_k$ & $t \times p_k$ (64$\times$2) precoding: $V^*_kW_k = I, V^*_i W_k = 0$ for $i \neq k$ \\
$A_k$ & $p_k \times q_k$ (4$\times$2) transmitting matrix, i.e. $A_k = H^*_kW_k$ \\
$n_k$ & $q_k$ (4) external noise\\
$n', n'', n'''$ & $q_k$ (4) normalized independent noise vectors \\
$x_k$ & $p_k$ (2)-symbol vector sent to UE $k$ \\
$y_k$ & $q_k$ (4)-symbol vector received by UE $k$ \\
$G_k$ & $p_k \times q_k$ (2$\times$4) detection matrix \\
$R_k$ & $q_k \times q_k$ (4$\times$4) noise covariance matrix \\
$L_k$ & $q_k \times q_k$ (4$\times$4) external noise power matrix, i.e., $n_k=L_kn'_k$ \\
$\Omega$ & $q_k \times q_k$ (4$\times$4) noiseless covariance matrix \\
\end{tabular}%
}
\caption{The algebraic notations used in the work and the typical matrix dimensions.}
\label{tab:notations}
\end{table}

To avoid confusion, we will choose $B_k$ (and, respectively, $V_k$) as \textit{maximal} reduced channels corresponding to some RCZF precoding. For example, usual zero-forcing matrices $W_k$ is obviously zero-forcing some non-trivial reduced channels $V'_k$, but we still choose $V_k$ corresponding to this RCZF to be $H_k$.
\par We are now ready to formulate the main result.
\begin{manualtheorem}{1}[\textbf{Interference Cancellation Condition}]\label{theorem:rczf}
1) In the noiseless MIMO system interference cancellation can be achieved only if precoding $W_k$ is reduced channel zero-forcing; 2) In this case, detection maps $G_k$ such that interference is canceled, exist; 3) These maps are unique, provided that $\mathop{\mathrm{rk}}B_k=\mathop{\mathrm{rk}}W_k$.
\end{manualtheorem}
The last condition is indeed practical, since RCZF condition $\mathop{\mathrm{rk}} V_kW_k = \mathop{\mathrm{rk}}W_k$ implies $\mathop{\mathrm{rk}}B_k\geqslant\mathop{\mathrm{rk}}W_k$, and we would usually like to choose $B_k$ as small as possible to satisfy less restrictions posed by $V_k W_j = 0$ for $j \neq k$ (these restrictions decrease effective transmission power).
\par
We will now see that Theorem~\ref{theorem:rczf} is surprisingly tautological. For the first statement, suppose that noiseless MIMO system cancels interference, and take $B_k=G_k$. Then RCZF conditions follow immediately from interference cancellation.
\par For the second statement, denote $\ker B_k = T_k \subset U_k$. Then, as  $V_k W_j = B_kH_kW_j = 0$ for $j \neq k$, we have $\ima H_kW_j \subset T_k$. Denote $\ima H_kW_k = P_k \subset U_k$. Since $\mathop{\mathrm{rk}} V_kW_k = \mathop{\mathrm{rk}} B_kH_kW_k = \mathop{\mathrm{rk}}W_k$, image of $H_kW_k$ is transverse to the kernel of $B_k$, i.e., $P_k \cap T_k = 0$. Now choose $G_k|_{T_k}=0, G_k|_{P_k}=(H_kW_k)^{-1}$ (this inverse is defined properly on $P_k$), and the rest of detection map arbitarily in case $P_k$ and $T_k$ do not generate $U_k$.
\par For the third statement, observe that subspaces $\ima H_k W_j \subset U_k$ generate the whole $T_k$. This is due to $B_k$ chosen in the maximal way --- otherwise we could replace it by $B'_k \supset B_k$ such that $T'_k = \ker B'_k \subset T_k$ is generated by $\ima H_k W_j$. Now, since $\mathop{\mathrm{rk}}B_k=\mathop{\mathrm{rk}}W_k$, we have $\dim T_k + \dim P_k = \dim U_k = q_k$, so $U_k = T_k \oplus P_k$, and $G_k$ is uniquely determined on both $P_k$ and $T_k$. Therefore, it is uniquely determined on $U_k$.

\section{Detection options}\label{sec:detection_options}

In this section we assume that precoding is RCZF and condition from the third statement of Theorem~\ref{theorem:rczf} is satisfied, i.e., $\mathop{\mathrm{rk}}B_k=\mathop{\mathrm{rk}}W_k$. Furthermore, we assume that $V_kW_k=I$ (which means that after establishing reduced channels $V_k$ precoding is obtained by taking pseudo-inverse, like in EZF). \par Existence and uniqueness of suitable detection maps $G_k$ established in the previous section is purely theoretical for now. In real-time MIMO systems user must be able to calculate $G_k$ from the information it is given (this information is provided by reference signals). To be more precise, we know neither all the maps participating in MIMO system, nor even those concerning us directly --- $H_k$ and all of $W_j$. What the $k$-th user does know (with decent precision), though, is transmitting matrix $A_k=H_kW_k$ and \textit{total covariance of the noise} given by the formula
$$R_k = H_k \left(\sum_{j \neq k} W_j W_j^*\right) H_k^* + L_kL_k^*,$$
where noise $n_k = L_k n'_k$ with $n'_k = \mathcal{N}(0,1)$. Sometimes $L_k$ is far from $I$ because this ``white'' noise may also contain inter-cell interference. Nevertheless, when we say that external noise tends to zero, we mean that $||L_k|| \to 0$. \par
In this notation one of widely used receivers, namely, MMSE-IRC detector, is given by the formula:
$$G_k = A_k^* (A_k A_k^* + R_k)^{-1}.$$
\begin{manualtheorem}{2}[\textbf{Interference Cancellation of MMSE-IRC}]\label{theorem:mmse_irc}
In the noiseless MIMO system with RCZF precoding MMSE-IRC detector coincides with the unique interference cancellation detector. In the noisy MIMO system, as $||L_k|| \to 0$, so does the restored signal.
\end{manualtheorem}
\begin{myproof}
For the first statement, we observe that $A_kA_k^* + R_k = H_k(\sum W_j W_j^*) H_k^*$. Denote this matrix by $\Omega$, so that $G_k=A_k^* \Omega^{-1}$. Though $\Omega$ depends on $H_k$, we do not add index to emphasize that inner sum is taken by all $k$. It is important to remark that, for $\Omega$ to be invertible, or, equivalently, $\mathop{\mathrm{rk}}\Omega=\dim U_k=q_k$, we require at least $q_k$ layers in total. Otherwise, as we can see from the formula, $\Omega$ is decomposed as a sequence of linear transformations passing through $< \! q_k$-dimensional vector space. Of course, this condition is almost always satisfied (for example, it is so whenever we have at least 2 users with 4 antennae and 2 layers each).
\par
We need to prove that, in the notation of previous section, $G_k = B_k$. Indeed,
$$A_k = H_kW_k = H_k \left(\sum W_k W_k^*\right)V_k^* = \Omega B_k^*,$$
since $W_k = W_k W_k^* V_k^*$, while $W_j W_j^* V_k = 0$ for $j \neq k$. Therefore, $G_k = (\Omega B_k^*)^* (\Omega)^{-1} = B_k$, since $\Omega$ is self-conjugate.
\par For the second statement, the final formula changes to $B_k \Omega (\Omega + L_kL_k^*)^{-1} = B_k (I + L_kL_k^*\Omega^{-1})^{-1}$, and the limit equals $B_k$ again.
\end{myproof}

We proceed to show that other widespread detection methods, namely, generalised LSE and QR-MLD, also cancel interference in MIMO system with RCZF precoding when the external noise power tends to zero. In particular, the limit of all these detecting techniques is the unique interference cancellation obtained in Theorem~\ref{theorem:rczf}.

\begin{figure}
    \centering
    \includegraphics[width=1\linewidth]{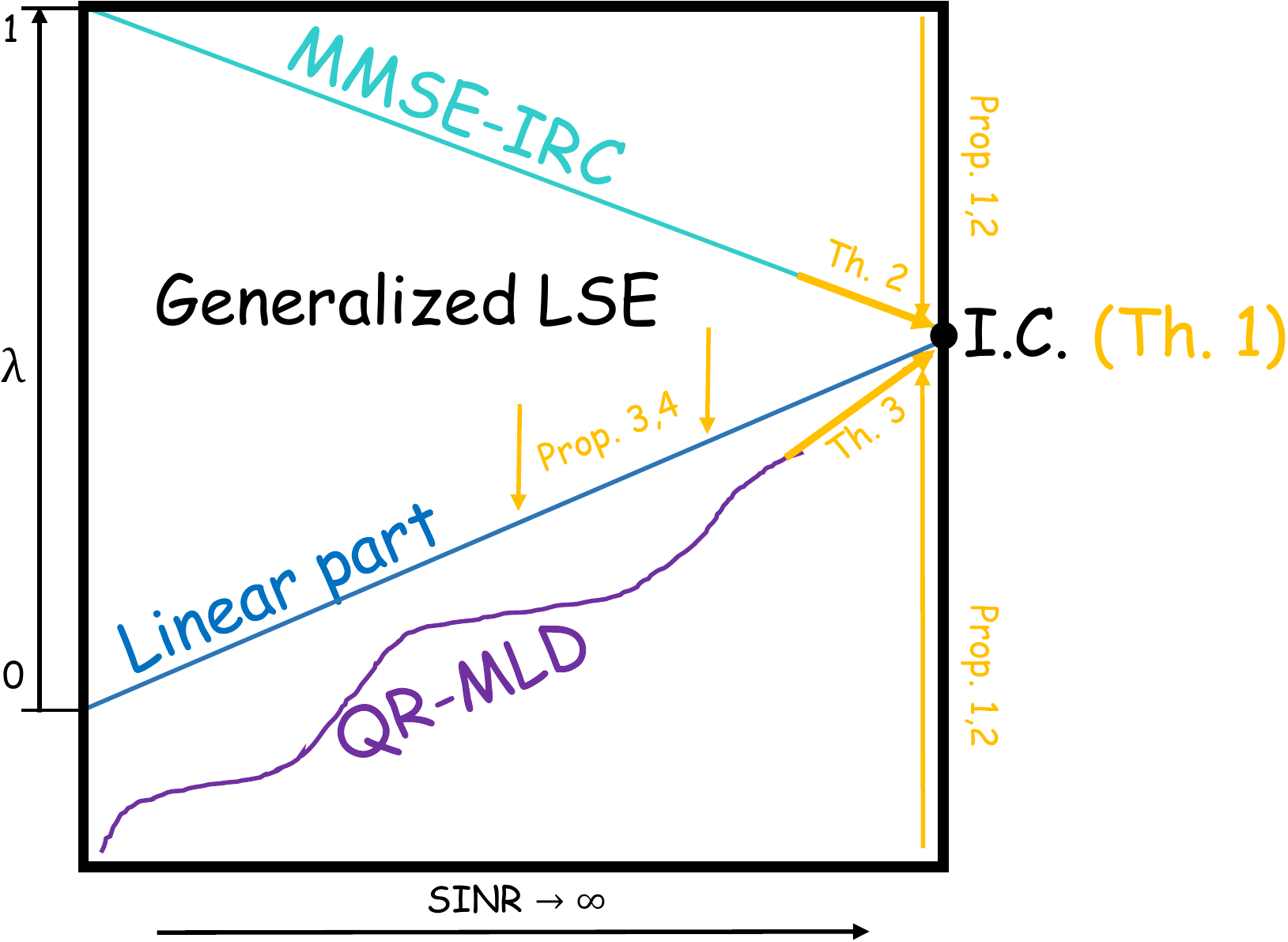}
    \caption{Detection limits and interference cancellation.}
    \label{fig:structure}
\end{figure}

\begin{prop}\label{prop:hermitian}
Let $F$ and $G$ be two self-conjugate operators on Hermitian space $V$ satisfying $\ima F \cap \ima G = 0$ and $\ima F \oplus \ima G = V$. Then $(F+\lambda G)^{-1}|_{\ima F}$ does not depend on $\lambda \neq 0$.
\end{prop}
\begin{myproof}
Observe that $(F+\lambda G)(\ker G) = F(\ker G) = \ima F$. It follows from the fact that $\ker F \cap \ker G = 0$ and $\ker F \oplus \ker G = V$ which, in its turn, is the consequence of $\ima F \perp \ker F$ and $\ima G \perp \ker G$. This bijection $\varphi \colon \ker G \to \ima F$ clearly does not depend on $\lambda$. It follows that $(F+\lambda G)^{-1}|_{\ima F} = \varphi^{-1}$ also does not depend on $\lambda$.
\end{myproof} \par
\begin{prop}
In the noiseless MIMO system with RCZF precoding the generalized LSE detector $A^*(AA^* + \lambda R)^{-1}$ does not depend on $\lambda \neq 0$.
\end{prop}
\begin{myproof}
We have $\ima AA^* = \ima A = P_k \subset U_k$ as in Theorem~\ref{theorem:rczf}. Similarly, $\ima R = T_k \subset U_k$, as we saw in the proof of the third statement of Theorem~\ref{theorem:rczf}. Since $P_k \cap T_k = 0$ and $U_k = P_k \oplus T_k$, we are in conditions of Prop.~\ref{prop:hermitian}, and $(AA^* + \lambda R)^{-1}|_{\ima AA^*}$ does not depend on $\lambda$. Observe that $\ker AA^* \oplus \ima AA^* = U_k$ and $\ker AA^* = \ker A^*$. We have two projections $\pi_1 \colon U_k \to \ker AA^*$ and $\pi_2 \colon V \to \ima AA^*$, and for any $v \in U_k$ we have $\pi_1(v)+\pi_2(v)=v$. Decomposing
\begin{multline*}A^*(AA^* + \lambda R)^{-1}(v) = A^*(\pi_1 + \pi_2)(AA^* + \lambda R)^{-1}(v)= \\ = A^* \pi_2 (AA^* + \lambda R)^{-1}(v)  = A^*((AA^* + \lambda R)^{-1} \iota_2)^* (v) = \\ = A^*((AA^* + \lambda R)^{-1}|_{\ima AA^*})^*(v)
\end{multline*}
--- does not depend on $\lambda$. The third equality is true due to $(XY)^* = Y^*X^*$ and $\pi_2^* = \iota_2$, where $\iota_2 \colon \ima AA^* \to U_k$ is the embedding.  
\end{myproof} \par

Thus, we can see that generalized LSE detector also cancels interference in the limit.
\par We will now describe QR-MLD process. Its main feature is being symbol-wise, adjusting itself during intermediate steps (and therefore non-linear), but we will see that it is irrelevant in the consideration of interference.

First, the QR-MLD finds square matrix $\mathcal{L}_k$ such that $\mathcal{L}\mathcal{L}^* = R_k$ (Cholesky decomposition, for instance, allows to do just that, though it is worth noting that matrix $\mathcal{L}$ with this property is not unique and may be replaced by any $\mathcal{L} U$ with $U$ unitary). Then the received vector in $U_k$ may be rewritten as $H_kW_k s_k + \mathcal{L} n''$, where $n''$ is normalized and independent noise.

The next and the main step of QR-MLD is to perform the QR-decomposition of matrix $\mathcal{L}^{-1} H_k W_k = Q \mathcal{R}$ (here $Q$ is unitary, and $\mathcal{R}$ is upper-triangular). It is at this moment when we face the main complication, namely, if $R_k$ does not contain external noise part $L_kL_k^*$ at all, then its rank is strictly less than $q_k$, and there is no way we can use $\mathcal{L}^{-1}$. Unable to deal with simplified scenario, we introduce small $L_k$ to make $\mathcal{L}$ invertible and study its behaviour when $||L_k||\to \! 0$ as previously. After establishing $Q$ and $\mathcal{R}$, QR-MLD detector extracts the diagonal part $\mathcal{R}_d$ of $\mathcal{R} = \mathcal{R}_d + \mathcal{R}_u$ and uses $\mathcal{R}^{-1}_d Q^* \mathcal{L}^{-1}$ as a first approximation of $G_k$. Applying this matrix to the received vector, we get:
\begin{multline*}
\mathcal{R}^{-1}_d Q^* \mathcal{L}^{-1} (\mathcal{L}Q\mathcal{R}s_k + \mathcal{L}n'') = \\ = \mathcal{R}^{-1}_d (\mathcal{R}_d +\mathcal{R}_u)s_k + \mathcal{R}^{-1}_d n''' = s_k + \mathcal{R}^{-1}_d \mathcal{R}_u s_k + \mathcal{R}^{-1}_d n''',
\end{multline*}

where $n'''=Q^*n''$ is again normalized independent noise. The second summand applies \textit{strictly} upper-triangular matrix $\mathcal{R}^{-1}_d \mathcal{R}_u$ to $s_k$ and, thus, we may unveil $s_k$ symbol by symbol starting from the end and subtracting the upper-triangular term. The third summand $\mathcal{R}^{-1}_d n'''$ which represents the effective noise after detection. \par
To conclude, the non-linearity of QR-MLD detection comes from symbol-wise discretization of decoded vector which is then plugged in the formula to find the rest of the symbols. It means that we can speak of \textit{linear part} of QR-MLD detection, which is equal to $\mathcal{R}^{-1}Q^*\mathcal{L}^{-1}$, and the difference between actual QR-MLD and its linear part tends to zero if the error itself tends to zero. We are now ready to prove Theorem~\ref{theorem:qr_mld_2}. \par
\begin{prop}\label{prop:generalized_lse}
In the conditions as above, when $\lambda \to 0$, generalized LSE detector $A^*(AA^* + \lambda R)^{-1} \to (A^*R^{-1}A)^{-1}A^*R^{-1}$.
\end{prop}
\begin{myproof}
Using matrix identities, obtain that $A^*(AA^* + \lambda R)^{-1} = (A^*R^{-1}A + \lambda I)^{-1}A^*R^{-1}$ and take the limit.
\end{myproof} \par
\begin{prop}
The linear part of the QR-MLD detector is equal to $(A^*R^{-1}A)^{-1}A^*R^{-1}$, provided that external noise is non-zero (so that $R$ is invertible).
\end{prop}
\begin{myproof}
We have $Q\mathcal{R}=\mathcal{L}^{-1}A \Leftrightarrow A = \mathcal{L}Q\mathcal{R}$, where $R = \mathcal{L}\mathcal{L}^*$. Substituting this into $(A^*R^{-1}A)^{-1}A^*R^{-1}$, we get $\mathcal{R}^{-1}Q^*\mathcal{L}^{-1}$ which is equal to the linear part of QR-MLD detector.
\end{myproof} \par
\begin{manualtheorem}{3}[\textbf{Interference Cancellation of QR-MLD}]\label{theorem:qr_mld_2}
In the noisy MIMO system with RCZF precoding, as $||L_k|| \to 0$, the limit of QR-MLD detector coincides with the unique interference cancellation detector.
\end{manualtheorem}
\begin{myproof}
Substituting $\lambda = 1$ in formula $A^*(AA^* + \lambda R)^{-1}$, we get MMSE-IRC detector which eliminates interference with zero noise as was shown in Theorem~\ref{theorem:mmse_irc}. By Prop.~2 this is also true for arbitrarily small $\lambda$. By Prop.~3 and Prop.~4, this is also what QR-MLD achieves.
\end{myproof}

The structure of our detection investigations is summarized in Fig.~\ref{fig:structure}.

\section{Experiments}\label{sec:experiments}

\begin{figure}
    \centering
    \includegraphics[width=1\linewidth]{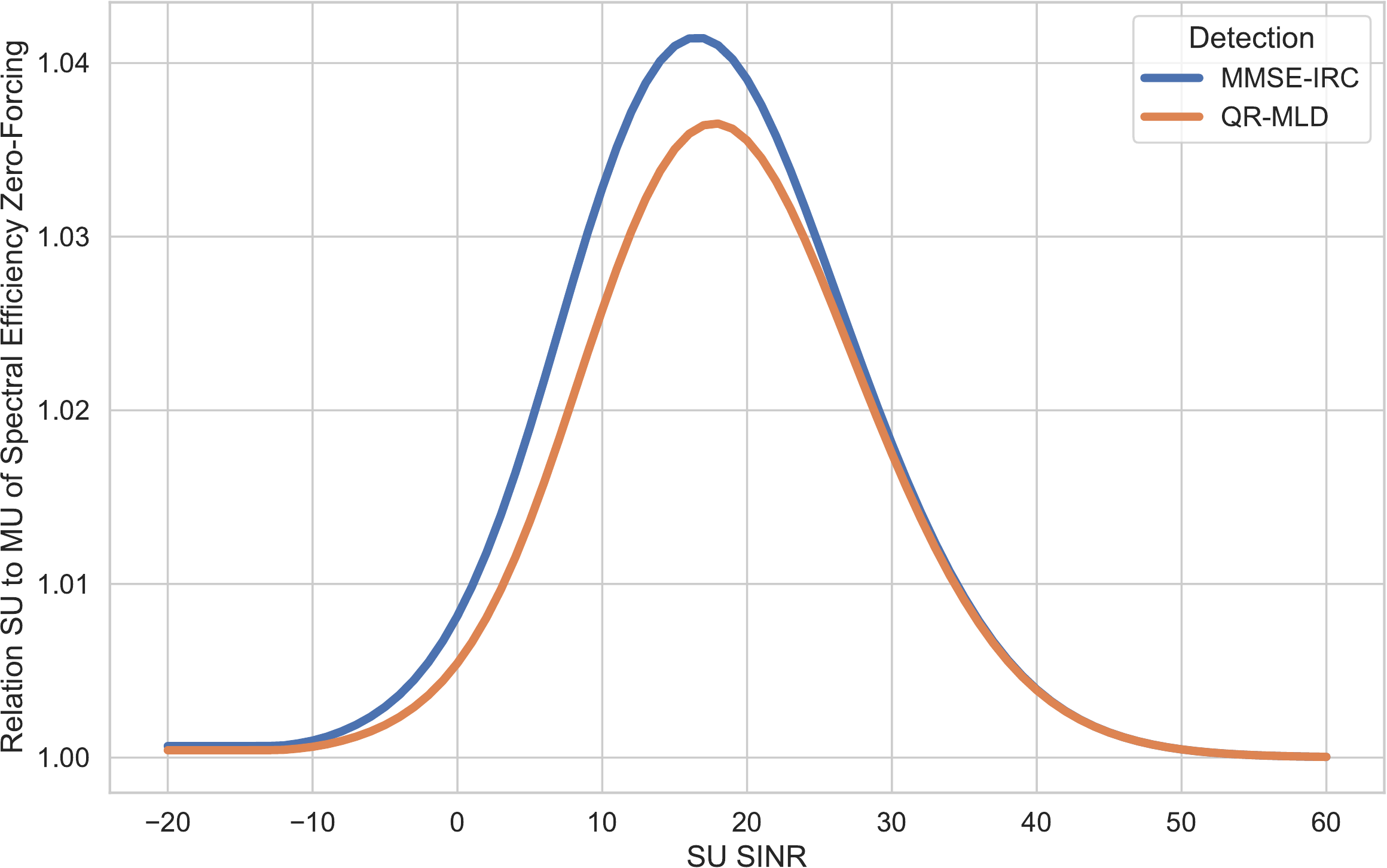}
    \caption{Ratio of Single-User to Multi-User Spectral Efficiency for MMSE IRC and QR-MLD detection matrices assuming Zero-Forcing precoding.}
    \label{Ratio SU MU}
\end{figure}

To support theoretical research, we conduct several important experiments. The goal to these experiments is to show the sufficiency and necessity of  Theorem~\ref{theorem:rczf} on practical examples. We used the Quadriga simulator~\cite{Quadriga} to create channel matrices. We draw all figures depending on average user Single-User (SU) SINR (db). Firstly, we study interference cancellation property of the QR-MLD and MMSE-IRC detection matrices and Zero-Foring (ZF) precoding matrix in Fig.~\ref{Ratio SU MU}. This figure plainly demonstrates interference cancellation for precoding of Reduced Channel Zero-Forcing (RCZF) class~\eqref{def:RCZF}. It shows the ratio between Single-User (SU) and Multi-User (MU) Spectral Efficiency. And therefore, by definition, when this ratio approaches unity, interference between users is canceled. The successful inter-user interference cancellation using the proper precoding and detection matrices experimentally proves the sufficiency of Theorem~\ref{theorem:rczf}.

\begin{figure}
    \centering
    \includegraphics[width=1\linewidth]{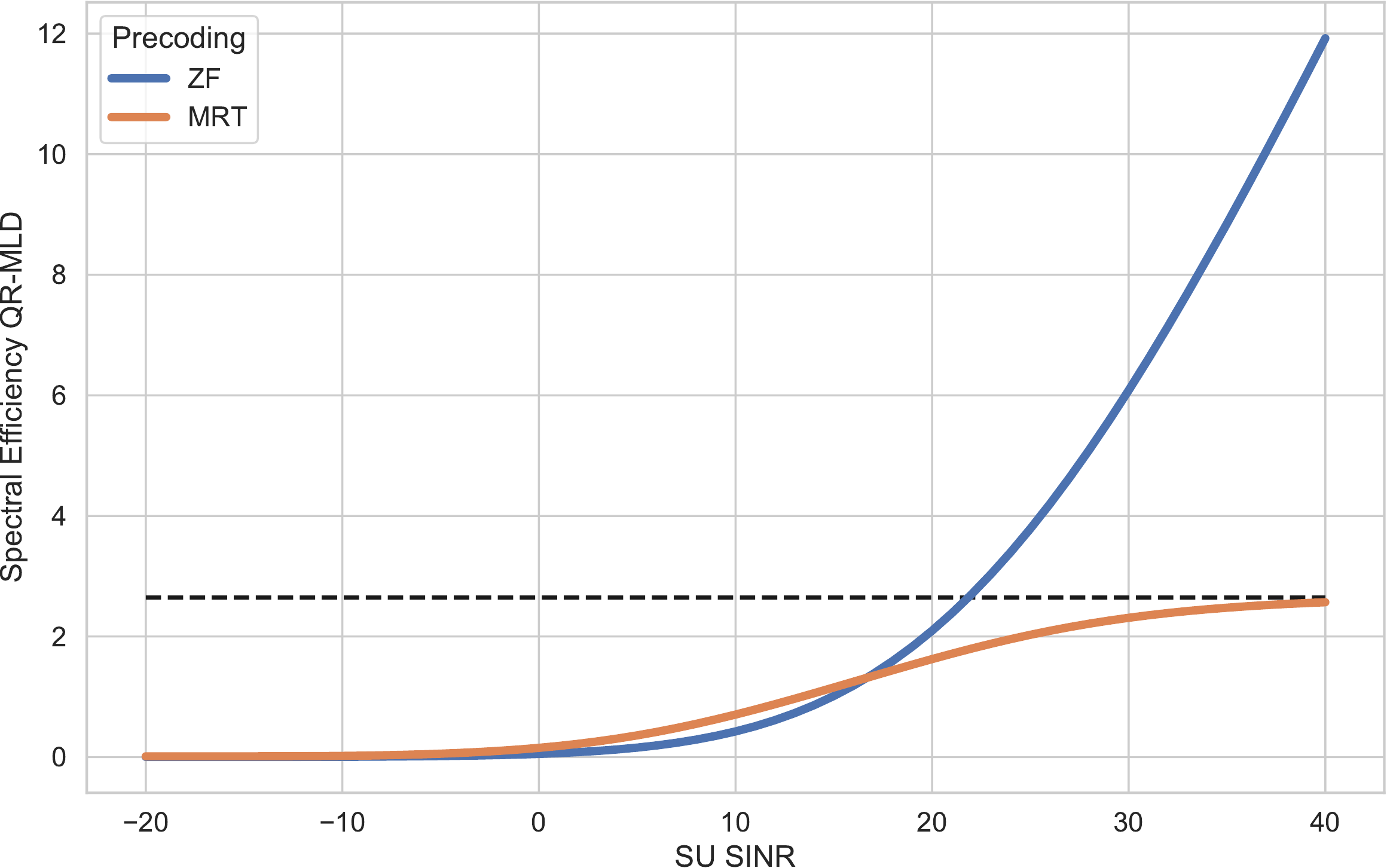}
    \caption{Multi-User Spectral Efficiency for QR-MLD detection matrix assuming Zero-Forcing (ZF) and Maximum Ratio Transmission (MRT) precodings matrices.}
    \label{MU Bad Precoding}
\end{figure}

To show experimental evidence of the necessity, we investigate the case when the conditions of Theorem~\ref{theorem:rczf} are violated. In Fig.~\ref{MU Bad Precoding} we take the Maximum Ratio Transmission (MRT) as a precoding matrix out of the RCZF class~\eqref{def:RCZF}. We added ZF precoding from the RCZF class for comparison with MRT. Additionally, QR-MLD was chosen as a detection matrix. In Fig.~\ref{MU Bad Precoding} we study the dependence of MU SE on SU SINR value. The value of MU SE grows using the ZF precoding when average SU SINR tends to infinity. In this case, the system only encounters negligible white noise. Unlike it, precoding that does not belong to the RCZF~\eqref{def:RCZF} class (we took MRT for instance), does not eliminate interference. The power of interference increases in this case. In Fig.~\ref{MU Bad Precoding}, we can see that MU SINR for MRT precoding approximates its limit denoted by black dotted asymptote.
%~\ref{Ratio SU MU Bad Precoding},
%In Fig.~\ref{Ratio SU MU Bad Precoding} we study the ratio between SU and MU SE. Interference is not canceled for MRT algorithm since the SU-MU ratio does not approach unity.

\begin{figure}
    \centering
    \includegraphics[width=1\linewidth]{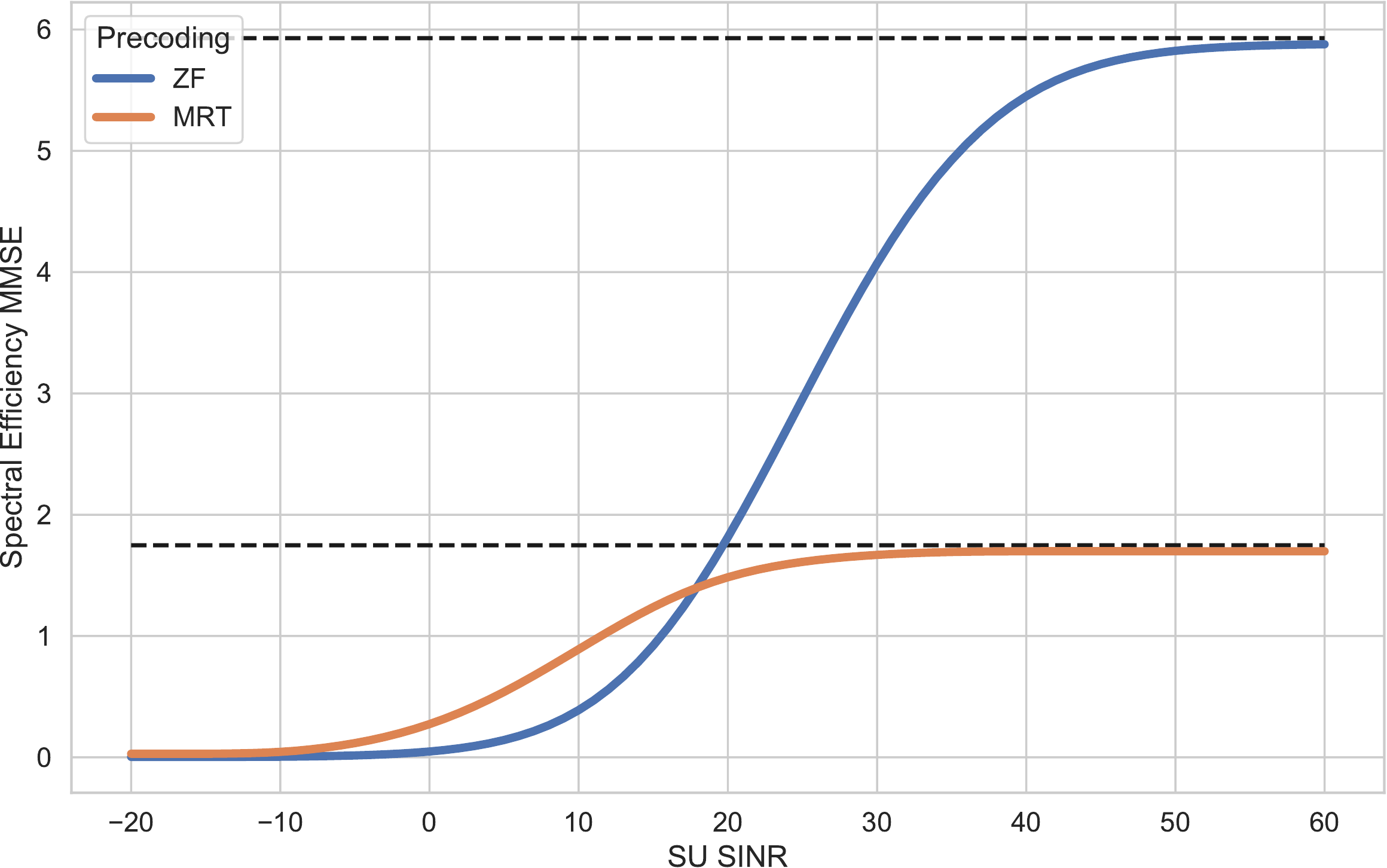}
    \caption{Multi-User Spectral Efficiency for MMSE detection matrix assuming Zero-Forcing (ZF) and Maximum Ratio Transmission (MRT) precoding matrices.}
    \label{MU MMSE}
\end{figure}

Limit of spectral efficiency also appears in the last comparison in Fig~\ref{MU MMSE}. In this experiment, the usual MMSE receiver is used instead of its MMSE-IRC counterpart. In conclusion, we would like to add that these experiments support the statement that only the combination of both precoding from the RCZF class~\eqref{def:RCZF} and generalized LSE detection (cf. Prop.~\ref{prop:generalized_lse} and Fig.~\ref{fig:structure}) class yields complete interference cancellation.
 
%The next experiment in Fig.~\ref{Ratio SU MU MMSE} also shows the situation of the violation of Theorem~\ref{theorem:rczf}. In this case, we take usual MMSE detection matrix, which assumes scalar noise-covariance matrix. For this experiment, we take ZF and MRT as the precoding matrices. As usual MMSE is not a generalized LSE detection matrix (cf. Prop.~\ref{prop:generalized_lse}) for the conditions of Theorem~\ref{theorem:rczf}, inter-user interference is not canceled. The ratio between SU and MU SE in this case tends to one.

\section{Conclusion}\label{sec:conclusion}

The bulk of results discussed in this paper is demonstrated in Fig.2 and can be summarized as follows. As SU SINR tends to infinity, there exist a unique linear detection cancelling inter-user interference in noisy MIMO scenario. This limit can be achieved by combining RCZF precoding with any receivers belonging to the family of generalized LSE detectors, as well as QR-MLD detection. These techniques exhaust all the possibilities for interference cancellation in MIMO systems.

%\medskip \par
%We will now reformulate conditions of RCZF in equivalent way. After this reformulation the statement and the proof of %Theorem~\ref{theorem:rczf} become almost tautological.
%\begin{prop} \label{equivRCZF}
%Set of precoding maps $W_i$ is RCZF if and only if there exist vector subspaces $T_i \subset R_i$ of receiver spaces such that $\ima %H_iW_i \cap T_i = 0$, and $\ima H_iW_j \subset T_i$ for $i \neq j$.
%\end{prop}
%\begin{myproof}
%Suppose $W_i$ is RCZF. Take $T_i = \ker B_i \subset R_i$. Since $V_iW_j=B_iH_iW_j=0$, we have $\ima H_iW_j \subset T_i$ for $i \neq j$. %Since $\mathop{\mathrm{rk}} V_iW_i = \mathop{\mathrm{rk}} B_iH_iW_i = \mathop{\mathrm{rk}}W_i$, we have $\ima H_iW_i \cap T_i = 0$. %\par
%Suppose that there exist subspaces $T_i$ satisfying conditions of Proposition~\ref{equivRCZF}. Similarly, take any $B_i$ such that %$\ker B_i = T_i$. One observes immediately that RCZF conditions hold for such $B_i$.
%\end{myproof} \par

% \begin{figure}
%     \centering
%     \includegraphics[width=1\linewidth]{Ratio SU to MU of Spectral Efficiency Bad Precoding} 
%     \caption{Ratio of Single-User to Multi-User Spectral Efficiency for QR-MLD detection matrix assuming Zero-Forcing (ZF) and Maximum Ratio Transmission (MRT) precoding matrices.}
%     \label{Ratio SU MU Bad Precoding}
% \end{figure}

% \begin{figure}
%     \centering
%     \includegraphics[width=1\linewidth]{Ratio SU to MU of Spectral Efficiency MMSE}
%     \caption{Ratio of Single-User to Multi-User Spectral Efficiency for MMSE detection matrix assuming Zero-Forcing (ZF) and Maximum Ratio Transmission (MRT) precoding matrices.}
%     \label{Ratio SU MU MMSE}
% \end{figure}

\bibliographystyle{IEEEtran}
\bibliography{references} 

\end{document}